\documentclass[twocolumn,showpacs,preprintnumbers,amsmath,amssymb]{revtex4}
\usepackage{color}
\usepackage{bm, graphicx, amsmath}
\begin{document}

\title{Maximum quantum violations of a class of Bell inequalities}
\author{V. U\u{g}ur G\"{u}ney and Mark Hillery}
\affiliation{Department of Physics, Hunter College of the City University of New York, 695 Park Avenue, New York, NY 10065 USA}

\begin{abstract}
We study a class of Bell inequalities and find their maximum quantum violation.  These inequalities involve $n$ parties, two measurements per party, with each measurement having two outcomes.  The $n=2$ case corresponds to the CH inequality.  We use the method of Jordan bases to find the maximum quantum violations.  Results are found for the cases $n=2$ through $n=7$.
\end{abstract}


\maketitle

\section{Introduction}
Bell inequalities are conditions that must be satisfied by any local, realistic theory \cite{bell}.  The predictions of quantum mechanics, which is not a local, realistic theory, can violate these inequalities.  Therefore, experiments that test these inequalities are able to tell us whether we need quantum mechanics to describe the world, or whether something less radical, such as a local realistic theory will do.  Experiments to date, have all shown that a local, realistic description is inadequate.  In addition, Bell inequalities have found applications in quantum cryptography, where they can be used to detect the presence of eavesdroppers \cite{ekert}.

Bell inequalities come in many varieties.  They can be characterized by the number of parties making measurements, $n$, the number of measurement settings, $p$, and the number of possible outcomes for each measurement, $m$.  The original versions, the CHSH \cite{chsh} and CH \cite{ch} inequalities were for the case $n=p=m=2$.   Kaszlikowsi, et al.\ showed that by increasing the number of outcomes, $p$, one could more strongly violate local realism \cite{kaszlikowski}.  Bell inequalities for the case $n=2$, $p=2$, and general $m$ were developed by Collins, et al.\ \cite{collins}, and this was generalized to the case of general $n$, $p=2$, and general $m$ by Son, et.\ al. \cite{son}. 

Though the subject initially received little attention, the literature on Bell inequalities is now extensive, and we cannot hope to summarize all of it here.  We would like to point out a very nice discussion by Nicolas Gisin, which summarizes what was known and what was not as of 2007 \cite{gisin}.  We would also like to mention approaches to Bell inequalities based on nonlocal games \cite{cleve}, graph theory \cite{cabello}, and logical consistency conditions \cite{abramsky}.

It is useful to know the largest violation of a given Bell inequality that quantum mechanics makes possible, and, in addition, which quantum state will produce this violation.  The original work on this subject was done by B.\ Cirel'son, who showed that while the bound for the correlations in the CHSH inequality dictated by local realism is $2$, the bound allowed by quantum mechanics is $2\sqrt{2}$ \cite{cirelson}.  Quantum bounds for other Bell inequalities have been found.  Wehner found bounds for Bell inequalities of the CHSH type for $n=2$, $m=2$, and general $p$ \cite{wehner}.  Pal and Vertesi combined numerical methods with analytic upper bounds to find quantum bounds for a large number of Bell inequalities with $n=2$, general $p$, and $m=2$. \cite{pal}  Here we will be considering inequalities belonging to the case $p=m=2$ with the object of determining their maximum quantum violation.  Full correlation Bell inequalities of this type were fully characterized by Werner and Wolf, and, in addition, they were able to show how to derive maximum quantum violations of these inequalities \cite{werner}.

The inequalities we study are not full correlation inequalities, and we will use a different method to find their maximum violation.  The class of inequalities we shall study can be considered generalizations of the CH inequality.  The technique we shall apply is that of Jordan bases \cite{jordan}.  Jordan bases are orthonormal bases of two subspaces, one for each subspace, with very nice overlap properties between basis elements corresponding to different spaces.  These bases can be constructed for any two subspaces and their properties allow us to place bounds on the eigenvalues of Bell operators, which do not depend on the dimension of the overall Hilbert space or of the subspaces.

\section{Inequalities}
We begin by deriving the inequalities we wish to consider.  Let us do it first for two parties, and then generalize to the case of $n$ parties.  Suppose that Alice can measure one of two observables, $a_{1}$ and $a_{2}$, and Bob can measure $b_{1}$ and $b_{2}$.  Each of these variables has two measurement outcomes, $\pm 1$.  Alice and Bob are presumed to be sufficiently far apart that their measurements are independent, i.e.\ Bob's choice of measurement will not influence the result of Alice's, and vice versa. A source produces two particles, and sends one to Alice and one to Bob, and then Alice and Bob each perform a measurement on their respective particle.  Local realism implies that the probabilities describing the measurement results satisfy
\begin{eqnarray}
\label{twoparty}
P(a_{1}=1,b_{1}=1)-P(a_{2}=1,b_{2}=1) \nonumber \\
\leq P(a_{2}=-1,b_{1}=1) + P(a_{1}=1,b_{2}=-1) .
\end{eqnarray}
Here, $P(a_{j}=m,b_{k}=m^{\prime})$ is just the probability that if Alice measures $a_{j}$ she gets $m$, and if Bob measures $b_{k}$ he gets $m^{\prime}$.  We shall prove a more general version of this inequality shortly.  This inequality is just a version of the CH Bell inequality.  A three party version, where we add another participant, Charlie, with observables $c_{1}$ and $c_{2}$ is
\begin{eqnarray}
\label{threeparty}
P(a_{1}=1,b_{1}=1,c_{1}=1)-P(a_{2}=1,b_{2}=1,c_{2}=1)  \leq  \nonumber \\
P(a_{2}=-1,b_{1}=1,c_{1}=1) + P(a_{1}=1,b_{2}=-1,c_{1}=1) \nonumber \\
 + P(a_{1}=1, b_{1}=1, c_{2}=-1) .
\end{eqnarray}
This can be extended to $n$ parties in the obvious way.

Now let us prove the $n$-party version.  In that case we label the observables as $a_{lj}$, where $l=1,2\ldots n$ and $j=1,2$, so each of the $n$ parties has two measurement choices, and each observable can take the values $\pm1$.  The inequality can be expressed as
\begin{eqnarray}
\label{nparty}
P(a_{11}=1,a_{21}=1, \ldots a_{n1}=1)  \leq \hspace{2cm} \nonumber \\
 P(a_{12}=1,a_{22}=1, \ldots a_{n2}=1) \nonumber \\ 
+ P(a_{12}=-1,a_{21}=1, \ldots a_{n1}=1) \nonumber \\
+ \ldots + P(a_{11}=1,a_{21}=1, \ldots a_{n-1,1}=1, a_{n2}=-1) .
\end{eqnarray}
Now if our experiment can be described by a local realistic theory, then there is a joint distribution for all of the observables, $P(a_{11},a_{21} \ldots a_{n1}; a_{12}, a_{22}, \ldots a_{n2})$, and all of the probabilities in the above inequality can be expressed in terms of this joint distribution.  The joint distribution gives the probability of sequences of ones and minus ones of length $2n$.  In order to show that this inequality is true, all we need to show is that every sequence of length $2n$ that appears in the probability on the left-hand side also appears in one of the probabilities on the right-hand side.  All of the sequences that appear on the left-hand side have their first $n$ elements equal to $1$, so these are the only sequences we need to consider.  If the sequence has all of its elements equal to $1$, then it contributes to the first probability on the right-hand side.  Now suppose that some of its elements in its second half (in the second set of $n$ elements) are equal to $-1$, and suppose that one of the $-1$'s corresponds to the observable $a_{l2}$.  This sequence will contribute to the probability $P(a_{11}=1, \ldots a_{l2}=-1,\ldots a_{n1}=1)$ on the right-hand side.  Therefore, all sequences that contribute to the probability on the left-hand side do contribute to a probability on the right-hand side, so the inequality is proved.

As was stated in the Introduction, these inequalities are not full correlation inequalities of the type considered by Werner and Wolf \cite{werner}.  A full correlation inequality contains only expectation values of products of $n$ observables, one from each party, i.e.\ $\langle s_{1}s_{2} \ldots s_{n}\rangle$, where $s_{j}$ can be either $a_{1j}$ or $a_{2j}$.  Our $n=2$ inequality, Eq.\ (\ref{twoparty}), can be expressed in this form, but those for $3$ or more parties cannot.

\section{Maximum quantum violation for 2 parties}
In order to illustrate our approach to finding the maximum quantum violation, we will start with the simplest case of two parties.  This will reproduce known results, but it gives a simple example of how the method based on Jordan bases works.  We have four observables, $a_{j}$ and $b_{k}$, $j,k=1,2$, which are now operators on the Hilbert space $\mathcal{H}_{a}\otimes\mathcal{H}_{b}$, where $\mathcal{H}_{a}$ is the Hilbert space in which Alice's quantum states lie, and $\mathcal{H}_{b}$ is the Hilbert space in which Bob's states lie.  Our first problem in finding the maximum violation of Eq.\ (\ref{twoparty}), is that we do not know how we should choose the operators corresponding to Alice's and Bob's observables.  For the moment let us specify them by their spectral projections.  Let $Q_{aj}$, $j=1,2$, be the projection operator onto the subspace of $\mathcal{H}_{a}$ on which $a_{j}$ has the eigenvalue $1$, and, similarly, $Q_{bj}$ is the projection operator onto the subspace of $\mathcal{H}_{b}$ on which $b_{j}$ has the eigenvalue $1$.  The projection corresponding to the subspace on which $a_{j}$ has the eigenvalue $-1$ is $I_{a}-Q_{aj}$, where $I_{a}$ is the identity on $\mathcal{H}_{a}$, the projection onto the subspace of $\mathcal{H}_{b}$ on which $b_{j}$ has eigenvalue $-1$ is $I_{b}-Q_{bj}$.  Defining the operator
\begin{eqnarray}
B_{2} & = & Q_{a1}Q_{b1} -Q_{a2}Q_{b2} - (I_{a}-Q_{a2})Q_{b1} \nonumber \\
& &  -Q_{a1}(I_{b} - Q_{b2}) ,
\end{eqnarray}
the condition in Eq.\ (\ref{twoparty}) can be rephrased as $\langle B_{2}\rangle \leq 0$.  Our task is to find the largest positive eigenvalue of $B_{2}$, which will be the largest quantum violation of Eq.\ (\ref{twoparty}).

Now let $S_{a1}$ and $S_{a2}$ be the subspaces of $\mathcal{H}_{a}$ onto which $Q_{a1}$ and $Q_{a2}$ project, respectively.  These subspaces possess orthonormal bases, Jordan bases, with the following property.  Let $\{ |u_{1j}\rangle \}$ be an orthonormal basis for $S_{a1}$ and $\{ |u_{2k}\rangle \}$ be an orthonormal basis for $S_{a2}$.  Note that we do not know the dimension of either $S_{a1}$ or $S_{a2}$, so we cannot specify how many vectors are in each basis.  These bases can be chosen so that
\begin{equation}
\langle u_{1j}|u_{2k}\rangle = \delta_{jk}\cos \theta_{aj} ,
\end{equation}
where the $\{ \theta_{aj} \}$ are known as the Jordan angles and lie between $0$ and $\pi /2$.  Similarly, one can define subspaces $S_{b1}$ and $S_{b2}$ corresponding to the ranges of $Q_{b1}$ and $Q_{b2}$, respectively, and Jordan bases $\{ |v_{1j}\rangle \}$ and $\{ |v_{2k}\rangle \}$ satisfying 
\begin{equation}
\langle v_{1j}|v_{2k}\rangle = \delta_{jk}\cos \theta_{bj} ,
\end{equation}
where $\{ |v_{1j}\rangle \}$ is an orthonormal basis for $S_{b1}$, and $\{ |v_{2k}\rangle \}$ is an orthonormal basis for $S_{b2}$.  Note that the projection operators can be expressed as
\begin{eqnarray}
Q_{al} & = & \sum_{j} |u_{lj}\rangle\langle u_{lj}| \nonumber \\
Q_{bm} & = & \sum_{k} |v_{mk}\rangle\langle v_{mk}| ,
\end{eqnarray}
where $l,m=1,2$.

We now want to find the largest positive eigenvalue of $B_{2}$.  We first note that any eigenstate with a positive eigenvalue must contain terms for which the first term in $B_{2}$ is nonzero, as it is the only positive term, that is the only term that can have a positive expectation value.  Therefore, the eigenstate must contain terms of the form $|u_{1j}\rangle \otimes |v_{1j}\rangle$, so we need to find out what happens if $B_{2}$ acts on states of this form.  We shall actually be a bit more general and consider the behavior of the operator $B_{2}$ acting on states of the form $|u_{lj}\rangle \otimes |v_{mk}\rangle$, where $l,m=1,2$.  What we find is that only states with the same values of $j$ and $k$ are coupled, i.e.\ $B_{2}$ acting one one of these states yields a linear combination of states with the same values of $j$ and $k$.  In particular, if we have that
\begin{equation}
|\psi\rangle = \sum_{l,m=1}^{2}c_{lm} |u_{lj}\rangle \otimes |v_{mk}\rangle ,
\end{equation}
then we have that we can represent $B_{2}|\psi\rangle$ as
\begin{equation}
\label{two-party-matrix}
B_{2}|\psi\rangle = \left( \begin{array}{cccc} -1 & 0 & 0 & x_{a}x_{b} \\ x_{b} & 0 & x_{a}x_{b} & 0 \\ x_{a} & x_{a}x_{b} & 0 & 0 \\ -x_{a}x_{b} & -x_{a} & -x_{b} & -1 \end{array} \right) \left(\begin{array}{c} c_{11} \\ c_{12} \\ c_{21} \\ c_{22} \end{array}\right) ,
\end{equation}
where we have set $x_{a}=\cos\theta_{aj}$ and $x_{b}=\cos\theta_{bk}$.  What this means is that $\mathcal{H}_{a}\otimes\mathcal{H}_{b}$ splits up into four-dimensional invariant subspaces under the action of $B_{2}$, and if we want to find the eigenvalues of $B_{2}$, we can examine each four-dimensional subspace individually.  The characteristic equation of the above matrix is
\begin{equation}
\lambda^{2}(\lambda^{2}+1) -x_{a}^{2}(1-x_{a}^{2})x_{b}^{2}(1-x_{b}^{2}) = 0 ,
\end{equation}
yielding a maximum eigenvalue of 
\begin{equation}
\lambda_{max}= \frac{1}{2} \left\{ [ 1+ 4x_{a}(1-x_{a}^{2})^{1/2}x_{b}(1-x_{b}^{2})^{1/2}]^{1/2} -1 \right\} .
\end{equation}
We can now maximize $\lambda_{max}$ with respect to $x_{a}$ and $x_{b}$.  The maximum occurs when $x_{a}=x_{b}=1/\sqrt{2}$ giving a maximum value for $\lambda_{max}$ of $(\sqrt{2}-1)/2$.  This is the maximum quantum violation of the inequality in Eq.\ (\ref{twoparty}).

We can also find observables and a quantum state that attain this violation.  Note that in the above calculation all of the subspaces, $S_{al}$ and $S_{bl}$, for $l=1,2$ are one dimensional, and the overlap between the vectors in the $1$ and $2$ subspaces is $1/\sqrt{2}$.  We can,  therefore, choose $a_{1}=\sigma_{za}$, $a_{2}=\sigma_{xa}$, $b_{1}=\sigma_{zb}$, and $b_{2}=\sigma_{xb}$, where $\sigma_{za}$ is just the $\sigma_z$ Pauli matrix acting on the two-dimensional space $\mathcal{H}_{a}$, $\sigma_{zb}$ is the $\sigma_z$ matrix acting in the two-dimensional space $\mathcal{H}_{b}$, and similarly for $\sigma_{xa}$ and $\sigma_{xb}$.  We then have that
\begin{eqnarray}
|u_{11} \rangle = |0\rangle_{a} & |v_{11}\rangle = |0\rangle_{b} \nonumber \\
|u_{21}\rangle = |+x\rangle_{a} & |v_{21}\rangle = |+x\rangle_{b} ,
\end{eqnarray}
where $\sigma_{z}|0\rangle = |0\rangle$, $\sigma_{z}|1\rangle = -|1\rangle$, and $|\pm x\rangle = (|0\rangle \pm |1\rangle )/\sqrt{2}$.  The state that produces the maximum violation, which is just the eigenstate of the matrix in Eq.\ (\ref{two-party-matrix}) with $x_{a}=x_{b}=1/\sqrt{2}$,  corresponding to $\lambda_{max}$, is, in the $\{ |0\rangle , |1\rangle \}$ basis,
\begin{eqnarray}
|\psi_{2}\rangle & = & \frac{1}{2}\left(\frac{1}{2+\sqrt{2}}\right)^{1/2} [ (1+\sqrt{2})(|0\rangle_{a}|0\rangle_{b} -|1\rangle_{a}|1\rangle_{b}) \nonumber \\
 & & + |0\rangle_{a}|1\rangle_{b} + |1\rangle_{a}|0\rangle_{b} ] .
\end{eqnarray}

This state has interesting properties.  First, it is a maximally entangled state.  Next, suppose Alice and Bob independently decide to measure their part of the state in either the $z$ or $x$ basis.  They then announce which basis they used.  We want to maximize the chance that, no matter which basis choice they make, that once the basis choices are announced, each party can predict the other's measurement result.  For example, suppose both parties decide to measure the state in the $z$ basis, and suppose Alice got $1$, corresponding to the state $|0\rangle$ for her result.  Then with a probability of $(2+\sqrt{2})/4 \simeq 0.85$, Bob will also have gotten the result $1$.  Similarly, if the both measured in the $x$ basis, and Alice got $1$ corresponding to $|+x\rangle$, then the probability that Bob got $-1$, corresponding to $|-x\rangle$, is $(2+\sqrt{2})/4$.  For all of the basis choices the correspondences are
\begin{eqnarray}
zz & |0\rangle_{a}\leftrightarrow |0\rangle_{b} & |1\rangle_{a}\leftrightarrow |1\rangle_{b} \nonumber \\
zx & |0\rangle_{a} \leftrightarrow |+x\rangle_{b} & |1\rangle_{a} \leftrightarrow |-x\rangle_{b} \nonumber \\
xz & |+x\rangle_{a} \leftrightarrow |0\rangle_{b} & |-x\rangle_{a} \leftrightarrow |1\rangle_{b} \nonumber \\
xx & |+x\rangle_{a} \leftrightarrow |-x\rangle_{b} & |-x\rangle_{a} \leftrightarrow |+x\rangle_{b} .
\end{eqnarray}
The first column gives the basis choices, with Alice's choice first, and the next two columns give which of Alice's and Bob's measurement results correspond to each other.  In all cases, if either Alice or Bob gets the measurement result shown above, then the probability that the other party obtains the corresponding state is $(2+\sqrt{2})/4$.  Therefore, in this state, Alice's and Bob's measurement results are highly correlated independent of whether they measure in the $z$ or the $x$ basis. It is correlations of this type that allow a quantum strategy of the CHSH nonlocal game to be better than any classical strategy \cite{cleve}.

\section{Three parties}
Now let us look at the three party inequality, Eq.\ (\ref{threeparty}).  Before finding the maximum violation, which will ultimately require some numerical work,  we present some simpler cases of quantum states that violate the inequality.  

One possibility is to find a state that makes the probability $P(a_{1}=1,b_{1}=1, c_{1}=1)$ nonzero and all of the others zero. In order to do this, we have to specify a quantum mechanical system and the observables.  We shall suppose that the system consists of three qubits, and that the observables  labeled by $1$ correspond to $\sigma_{z}$ and those labeled by $2$ correspond to $\sigma_{x}$.  That is, $a_{1}=\sigma_{za}$, $a_{2}=\sigma_{xa}$, etc.  Define the subspace $S$ to be the span of the vectors $\{ |+x,+x,+x\rangle , |-x,0,0\rangle , |0,-x,0\rangle , |0,0,-x\rangle \}$, where in these states, the first slot is the state of qubit $a$, the second the state of qubit $b$, and the third of qubit $c$.  What we want is a vector that is orthogonal to $S$ and has a nonzero overlap with the state $|0,0,0\rangle$.  Such a state is
\begin{eqnarray}
|\psi^{\prime}_{3}\rangle & = & \frac{1}{2\sqrt{2}}(|0,0,0\rangle + |0,0,1\rangle + |0,1,0\rangle \nonumber \\
& & + |1,0,0\rangle - |0,1,1\rangle - |1,0,1\rangle - |1,1,0\rangle \nonumber \\
& & -|1,1,1\rangle ).
\end{eqnarray}
With this state, the left-hand side of the inequality in Eq.~(\ref{threeparty}) becomes $1/8$, and the right-hand side is zero.  As we shall see, this is by no means the largest violation we can obtain.

This state has the following correlation properties.  If all of the parties measure in the $z$ basis, or two of them measure in the $x$ basis and the remaining party measures in the $z$ basis, then the measurement results are uncorrelated.  That is, one party does not gain any information from his measurement about what the results of the other two measurements were.  The situation is different, however, if all of the parties measure in the $x$ basis or one measures in the $x$ basis and two measure in the $z$ basis.  In order to see what happens when everyone measures in the $x$ basis, we can express the state as
\begin{eqnarray}
|\psi^{\prime}_{3}\rangle & = & \frac{1}{2} ( |+x,+x,-x\rangle + |+x,-x,+x\rangle \nonumber \\
& & + |-x,+x,+x\rangle - |-x,-x,-x\rangle ) .
\end{eqnarray}
From this expression, we see that if one party gets $+x$ for his measurement, he can be assured that the remaining parties got opposite results for theirs, that is one got $+x$ and the other got $-x$.  If one party gets $-x$, however, then he knows that the remaining parties got the same result for their measurements, either both got $+x$, or both got $-x$.  Now suppose that one party, say the first, measures in the $x$ basis while the other two measure in the $z$ basis.  It is now convenient to express the state as
\begin{eqnarray}
|\psi^{\prime}_{3}\rangle & = & \frac{1}{2}[ |+x\rangle (|0,0\rangle - |1,1\rangle ) \nonumber \\
& & + |-x\rangle ( |0,1\rangle + |1,0\rangle )] .
\end{eqnarray}
From this we see the following,  If the party measuring in the $x$ basis gets $+x$, then the other two parties will get the same measurement result, and if he gets $-x$, then the other two parties will get opposite results.  Now, if one of the parties measuring in the $z$ basis gets $0$, then the other two parties got either  $+x$ and $0$ or $-x$ and $1$, while if the $z$ result was $1$, then the other two parties got either $+x$ and $1$ or $-x$ and $0$.

One can obtain a larger violation of the inequality in Eq.~(\ref{threeparty}) with the same quantum system, three qubits, and the same assignment of observables, but choosing a different quantum state.  Under these assumptions the state that produces the largest violation is the eigenstate of the operator
\begin{eqnarray}
B_{3}^{\prime} & = & |0,0,0\rangle\langle 0,0,0| - |+x,+x,+x\rangle\langle +x,+x,+x| \nonumber \\
& & - ( |-x,0,0\rangle\langle -x,0,0| - |0,-x,0\rangle\langle 0,-x,0| \nonumber \\
& & - |0,0,-x\rangle\langle 0,0,-x| 
\end{eqnarray}
with the largest eigenvalue.  Setting the characteristic polynomial of the operator equal to zero, we find that there are three eigenvalues of zero, two of $-1/2$, and the remaining three eigenvalues are roots of the cubic equation, $8\lambda^{3}+16\lambda^{2} + 5\lambda -2=0$.  Solving this we find the one positive root is given, to three places, by $0.223$.  This is the largest violation of the inequality in Eq.~(\ref{threeparty}) with this choice of variables, and it represents an improvement over our previous value of $1/8$.  As we shall see, we can do better.

Let us now apply the method developed in the previous section.  We now make no assumptions as to what the observables are.  With the notation as before, the operator corresponding to the inequality in Eq.~(\ref{threeparty}) is
\begin{eqnarray}
B_{3} & = & Q_{a1}Q_{b1}Q_{c1} - Q_{a2}Q_{b2}Q_{c2} - (I_{a}-Q_{a2})Q_{b1}Q_{c1} \nonumber \\
& & - Q_{a1}(I_{b}-Q_{b2})Q_{c1} - Q_{a1}Q_{b1}(I_{c} - Q_{c2}) 
\end{eqnarray}
where the $Q_{lj}$ operators, $l\in \{a,b,c\}$ and $j=1,2$, are projections operators on one of the Hilbert spaces $\mathcal{H}_{a}$, $\mathcal{H}_{b}$, and $\mathcal{H}_{c}$.  We have the subspaces, $S_{lj}$, which are the ranges of the corresponding projections $Q_{lj}$ each with their Jordan bases.  In particular, $\{ |u_{1k}\rangle \}$ and $\{ |u_{2k}\rangle \}$ are the Jordan bases for $S_{a1}$ and $S_{a2}$, respectively, $\{ |v_{1k}\rangle \}$ and $\{ |v_{2k}\rangle \}$ are the Jordan bases for $S_{b1}$ and $S_{b2}$, respectively, and $\{ |w_{1k}\rangle \}$ and $\{ |w_{2k}\rangle \}$ are the Jordan bases for $S_{c1}$ and $S_{c2}$, respectively.  We now consider the action of $B_{3}$ on vectors of the form $|u_{lr}\rangle_{a}|v_{ms}\rangle_{b}|w_{nt}\rangle_{c}$, where $l,m,n=1,2$, and $r$, $s$, and $t$ are fixed.  We find that the subspace spanned by these eight vectors, which we shall call $T_{8,rst}$, is mapped into itself by $B_{3}$.  In fact, denoting 
\begin{equation}
\begin{array}{cc}
|\phi_{1}\rangle = |u_{1r}\rangle_{a}|v_{1s}\rangle_{b}|w_{1t}\rangle_{c} & |\phi_{4}\rangle = |u_{2r}\rangle_{a}|v_{1s}\rangle_{b}|w_{1t}\rangle_{c} \\ 
|\phi_{3}\rangle = |u_{1r}\rangle_{a}|v_{1s}\rangle_{b}|w_{2t}\rangle_{c} & |\phi_{5}\rangle = |u_{2r}\rangle_{a}|v_{2s}\rangle_{b}|w_{2t}\rangle_{c} \\
|\phi_{3}\rangle = |u_{1r}\rangle_{a}|v_{2s}\rangle_{b}|w_{1t}\rangle_{c} ,& 
\end{array}
\end{equation}
and defining $T_{5,rst}$ to be the five-dimensional subspace spanned by these vectors, we find that $B_{3}$ maps $T_{8,rst}$ into $T_{5,rst}$.  That means that if we wish to find nonzero eigenvalues of $B_{3}$, we only need to consider vectors in $T_{5,rst}$, which reduces our problem from an eight dimensional one to a five dimensional one.  In the subspace $T_{5,rst}$ and in the basis $\{ |\phi_{j}\rangle | j=1,2,\ldots 5 \}$, $B_{3}$ can be represented by the matrix
\begin{equation}
\left( \begin{array}{ccccc}
-2 & -x_{c} & -x_{b} & -x_{a} & x_{a}x_{b}x_{c} \\ x_{c} & 0 & x_{b}x_{c} & x_{a}x_{c} & 0 \\
x_{b} & x_{b}x_{c} & 0 & x_{a}x_{b} & 0 \\ x_{a} & x_{a}x_{c} & x_{a}x_{b} & 0 & 0 \\
-x_{a}x_{b}x_{c} & -x_{a}x_{b} & x_{a}x_{c} & -x_{b}x_{c} & -1 \end{array} \right) .
\end{equation}
The characteristic equation of this matrix is
\begin{eqnarray}
\label{characteristic3}
\lambda^{5} + 3\lambda^{4} + (2+\gamma -\alpha + \beta )\lambda^{3} + (\gamma -\alpha +\beta ) \lambda^{2} \nonumber \\
+ (2\gamma -\alpha -3) \lambda + \beta (\alpha -2\beta -1) = 0 ,
\end{eqnarray}
where
\begin{equation}
\begin{array}{cc}
\alpha = (x_{a}x_{b})^{2}+(x_{a}x_{b})^{2}+(x_{b}x_{c})^{2} & \beta = (x_{a}x_{b}x_{c})^{2} \\
\gamma = x_{a}^{2}+ x_{b}^{2}+x_{c}^{2} & . \end{array}
\end{equation}
We find numerically that the largest root of Eq.\ (\ref{characteristic3}) is achieved when $x_{a}=x_{b}=x_{c}$.  An analytic argument for this condition is given in the appendix.  Setting $x_{b}$ and $x_{c}$ equal to $x_{a}$, we find that the characteristic equation can be expressed as
\begin{eqnarray}
(\lambda +x_{a}^{2})^{2} [ \lambda^{3}+ (3-2x_{a}^{2})\lambda^{2} + (2-3x_{a}^{2}+x_{a}^{6})\lambda
\nonumber \\
+3x_{a}^{6} - 2x_{a}^{8} -x_{a}^{2}] = 0 .
\end{eqnarray}
The cubic equation does have a real, positive root, and that is the one in which we are interested.  We find that its maximum value occurs when $x_{a}=[(\sqrt{5}-1)/2]^{1/2} \simeq 0.786$ and $\lambda_{max}=\sqrt{5}-2 \simeq 0.236$.  Note that this is a larger violation that we obtained when the parties measured in either the $z$ or the $x$ basis.

This solution also tells us what observables we should use to obtain the maximum violation, and gives us the state state that produces this violation.  For $a_{1}$, $b_{1}$, and $c_{1}$ we choose, as before, $\sigma_{z}$.  Next, define the orthonormal vectors
\begin{eqnarray}
|u_{+}\rangle & = & x_{a}|0\rangle + \sqrt{1-x_{a}^{2}} |1\rangle \nonumber \\
|u_{-}\rangle & = & - \sqrt{1-x_{a}^{2}} |0\rangle + x_{a} |1\rangle ,
\end{eqnarray}
and for the operators $a_{2}$, $b_{2}$, and $c_{2}$ we choose $|u_{+}\rangle\langle u_{+}| - |u_{-}\rangle\langle u_{-}|$.  The state that achieves the maximum violation with this choice of observables is
\begin{eqnarray}
|\psi_{3}\rangle & = & \left( 4-\frac{8}{\sqrt{5}}\right)^{1/2} |000\rangle + \left( -\frac{3}{2} + \frac{7}{2\sqrt{5}}\right)^{1/2} \nonumber \\
& & (-|001\rangle - |010\rangle - |100\rangle +|111\rangle ) \nonumber \\
& & - \left( 1-\frac{2}{\sqrt{5}}\right)^{1/2}( |011\rangle + |101\rangle + |110\rangle .
\end{eqnarray} 
Note that this is not a GHZ state, i.e.\ it cannot be transformed by local unitaries into a state of the form$(|000\rangle + |111\rangle )\sqrt{2}$.  If it could, when we formed a density matrix from the state and traced out Bob and Charlie, we would obtain a reduced density matrix proportional to the identity.  That does not happen with $|\psi_{3}\rangle$.  Werner and Wolf showed that for all full correlation Bell inequalities with two measurement settings per party and each measurement having two outcomes, the maximally violating states are $n$-party generalizations of GHZ states \cite{werner}.  The fact that $|\psi_{3}\rangle$ is not a GHZ state is a result of the fact that the inequality in Eq.\ (\ref{threeparty}) is not a full correlation inequality.  

Finally, we note that the situation described by the inequality in Eq.\ (\ref{threeparty}) and its violation can be described in terms of a nonlocal game \cite{cleve}.  Each of the three parties,  is sent an instruction bit by a referee, and they then send a bit back to the referee, who determines whether the parties have won the game or not.  The parties are not allowed to communicate once the game has started.  There are only five possible sets of instructions, and they are equally probable.  They are either all zero, all one, or one of them is one and the other two are zero. The conditions for winning are
\begin{enumerate}
\item  If all instruction bits are $0$, then each party must return a $0$.
\item  If all instruction bits are $1$, then not all parties return a $0$, i.e.\ they only lose if all of them return a $0$.
\item If two of the instruction bits are $0$ and the remaining one is $1$, then they only lose if the party who received a $1$ returns $1$ and the other two return $0$.
\end{enumerate}

Let us first consider a classical strategy.  The optimal classical strategy is a deterministic one in which the bit each party sends is a function of the instruction bit they receive \cite{cleve}.  We shall show that any classical strategy must fail for at least one of the sets of instruction bits, which means that the maximum probability of winning is $4/5$.  We shall then present a strategy that does succeed with this probability, which proves that this is the optimal classical probability of winning.  Now, suppose one of the parties receives an instruction bit of $0$.  Then in order to win in the case all of the instruction bits are $0$, each party must return a bit of $0$.  Now consider what happens when one of the parties receives a $1$.  If the other two parties receive a $0$, then they will return a $0$, so in order to win, the party we are considering should return a $0$.  So that means in all cases, each party should return a $0$.  However, if the instruction set consists of all $1$'s, then they will all return $0$'s and lose.  Consequently, they cannot win all of the time, so the maximum winning probability is $4/5$ and the strategy where all parties always return $0$ achieves this probability.   

Now let us consider a quantum strategy.  The parties share a quantum state, and if they receive an instruction bit $0$ they measure observable $1$ (that is, $a_{1}$, $b_{1}$, and $c_{1}$), and if they receive an instruction bit $1$, they measure observable $2$.  They then send a bit corresponding to their measurement result, if their result is $1$ they send $0$, and if their result is $-1$ they send $1$.  Their probability of winning, $p_{\text{quant}}$, is just
\begin{equation}
p_{\text{quant}}=\frac{4}{5} + \frac{1}{5}\Delta ,
\end{equation}
where $\Delta$ is given by the left-hand side of the inequality in Eq.\ (\ref{threeparty}) minus the right-hand side. If Alice, Bob, and Charlie share $|\psi_{3}\rangle$ and make the measurements that maximally violate the inequality in Eq.\ (\ref{threeparty}), then they achieve a winning probability of 
\begin{equation}
p_{\text{quant}}=\frac{4}{5} + \frac{1}{5}(\sqrt{5}-2) \simeq 0.8472 ,
\end{equation}
which is better than the classical result.

\section{More than three parties}
The same technique can be used to find maximum quantum violations of the inequality in Eq.\ (\ref{nparty}) for $n$ parties.  Each new party adds an additional Jordan angle.  In all cases we examined, we find, numerically, that the maximum value of the positive root of the characteristic equation is achieved when all of the Jordan angles are equal.  This further implied that the relevant eigenvalue is a root of the cubic equation (this has only been verified up to $n=7$)
\begin{eqnarray}
\lambda^{3}+ [n-(n-1)x_{a}^{2}]\lambda^{2} + (n-1-nx_{a}^{2} + x_{a}^{2n})\lambda \nonumber \\
+ nx_{a}^{2n}-(n-1)x_{a}^{2n+2}-x_{a}^{2}=0 .
\end{eqnarray}
The results are plotted in Figure 1, where we show the maximum quantum violation as a function of the number of parties.
\begin{figure}
\includegraphics[scale=.60]{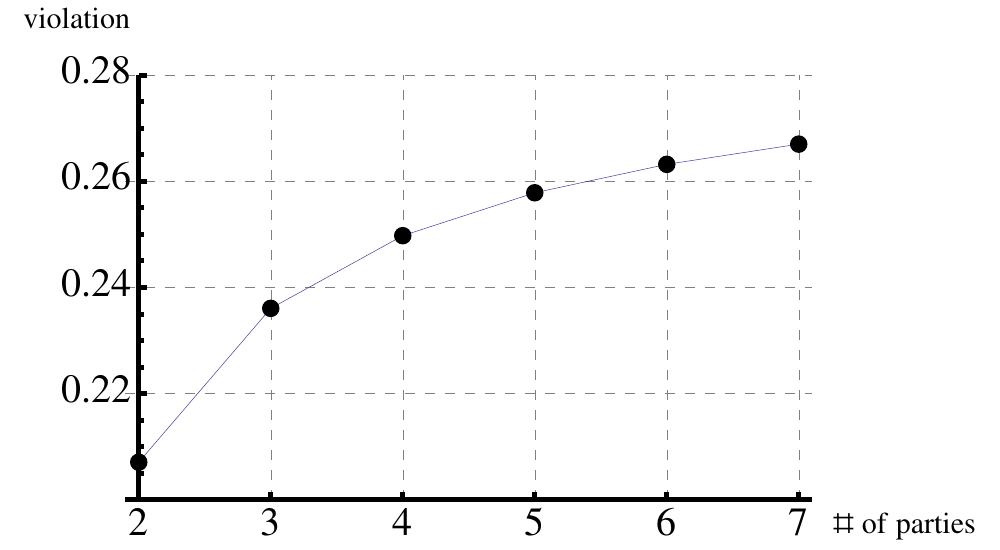}
\caption{Size of maximum quantum violation of inequality versus number of parties} 
\end{figure}
We also give the maximum violations, $\lambda_{max}$, and the values of $x_{a}$ that produce them in the following table: \newline

\begin{tabular}{|c|c|c|} \hline
$n$ & $x_{a}$ & $\lambda_{max}$ \\ \hline 3 & 0.786151 & 0.236068 \\ 4 & 0.830913 & 0.249757 \\
5 & 0.860012 & 0.257836 \\ 6 & 0.880509 & 0.263187 \\ 7 & 0.895745 & 0.266998 \\ \hline
\end{tabular}

\  \newline
\noindent
As can be seen from both the figure and the table, the size of the maximum violation increases with the number of parties.  The values of $x_{a}$ can be used to construct the observables that achieve the maximum violation.

\section{Conclusions}
We have shown how the technique of Jordan bases for two subspaces can be used to find maximum quantum violations of a class of Bell inequalities.  We do not need to make any assumptions on the dimension of the Hilbert space, and the technique gives us the observables and the states that produce the maximum violations.

\section*{Acknowledgments}
This research was partially supported by a PSC-CUNY grant.

\section*{Appendix}
We would like to show that if a positive root of Eq.\ (\ref{characteristic3}) reaches a maximum when $x_{a}=x_{b}=x_{c}$.  We first note that there is a positive root of this equation, because it is not too hard to show that when $\lambda = 0$, the value of the characteristic polynomial is negative, but its value is clearly positive for sufficiently large $\lambda$.  

We begin by defining 
\begin{eqnarray} 
F(\lambda , x_{a},x_{b},x_{c}) & = & \lambda^{5} + 3\lambda^{4} + (2+\gamma -\alpha + \beta )\lambda^{3} \nonumber \\
& & + (\gamma -\alpha +\beta ) \lambda^{2} + (2\gamma -\alpha -3) \lambda \nonumber \\
& & + \beta (\alpha -2\beta -1) .
\end{eqnarray}
The characteristic equation, $F(\lambda , x_{a},x_{b},x_{c}) =0$, now defines $\lambda$ as a function of $x_{a}$, $x_{b}$, and $x_{c}$.
We are interested in points where $\lambda (x_{a},x_{b},x_{c})$ is a maximum, which means that we want
\begin{equation}
\label{partials1}
\frac{\partial \lambda}{\partial x_{a}} = \frac{\partial \lambda}{\partial x_{b}} = \frac{\partial \lambda}{\partial x_{c}} = 0 .
\end{equation}
Now we have that 
\begin{equation}
\frac{\partial F}{\partial\lambda}\frac{\partial \lambda}{\partial x_{a}} + \left( \frac{\partial F}{\partial x_{a}}\right)_{\lambda} =0  ,
\end{equation}
and similarly for the derivatives with respect to $x_{b}$ and $x_{c}$.  The subscript on the second term indicates that $\lambda$ is held constant during this differentiation.  Therefore, the condition in Eq.\ (\ref{partials1}) becomes
\begin{equation}
\left( \frac{\partial F}{\partial x_{a}}\right)_{\lambda} = \left( \frac{\partial F}{\partial x_{b}}\right)_{\lambda} = \left( \frac{\partial F}{\partial x_{c}}\right)_{\lambda} = 0 .
\end{equation}
Defining the three functions of $x_{a}$, $x_{b}$, and $x_{c}$, 
\begin{eqnarray}
f_{0}=\beta (\alpha -2\beta -1) & \hspace{5mm} & f_{2}=\gamma -\alpha + \beta \nonumber \\
f_{1}=\beta (2\gamma - \alpha - 3) , & & 
\end{eqnarray}
the above equations become
\begin{eqnarray}
\frac{\partial f_{2}}{\partial x_{a}} (\lambda^{3}+\lambda^{2}) + \frac{\partial f_{1}}{\partial x_{a}} \lambda + \frac{\partial f_{0}}{\partial x_{a}} &  = & 0 \nonumber \\
\frac{\partial f_{2}}{\partial x_{b}} (\lambda^{3}+\lambda^{2}) + \frac{\partial f_{1}}{\partial x_{b}} \lambda + \frac{\partial f_{0}}{\partial x_{b}} & = & 0 \nonumber \\
\frac{\partial f_{2}}{\partial x_{c}} (\lambda^{3}+\lambda^{2}) + \frac{\partial f_{1}}{\partial x_{c}} \lambda + \frac{\partial f_{0}}{\partial x_{c}} & = & 0 .
\end{eqnarray}
These equations can be rearranged to eliminate the $\lambda^{3}$ and $\lambda^{2}$ terms, leaving us with three equations that are linear in $\lambda$.  For example, from the first two equations we find
\begin{eqnarray}
x_{b}(x_{a}^{2}-1)\left( \frac{\partial f_{1}}{\partial x_{a}} \lambda + \frac{\partial f_{0}}{\partial x_{a}}\right) \nonumber \\
= x_{a}(x_{b}^{2}-1)\left( \frac{\partial f_{1}}{\partial x_{b}} \lambda + \frac{\partial f_{0}}{\partial x_{b}}\right) ,
\end{eqnarray}
which becomes
\begin{eqnarray}
(x_{a}^{2}-x_{b}^{2}) [ f_{1} -x_{a}^{2}x_{b}^{2} \beta (x_{c}^{2}-1)]\lambda  \nonumber \\
+ (x_{a}^{2}-x_{b}^{2}) [ f_{0} -x_{a}^{2}x_{b}^{2} \beta (x_{c}^{2}-1)] =0 ,
\end{eqnarray}
The two remaining conditions can be found from the one above by exchanging $x_{b}$ and $x_{c}$ to obtain the second condition, and by exchanging $x_{a}$ and $x_{c}$ to obtain the third (note that $f_{0}$, $f_{1}$, and $f_{2}$ are symmetric in $x_{a}$, $x_{b}$, and $x_{c}$).  Now, if $x_{a}=x_{b}=x_{c}$, these equations will be satisfied.  If we assume that any two of the $x_{a}$, $x_{b}$, and $x_{c}$ are all different, then we find that it is necessary that $f_{0}=f_{1}$.  This, however, implies that $\lambda =-1$, which is not a positive root.  Therefore, if a positive root is to have a maximum, then we need $x_{a}$, $x_{b}$, and $x_{c}$ to be equal.


\begin{thebibliography}{99}
\bibitem {bell} J.~S.~Bell, Physics {\bf 1}, 195 (1964).
\bibitem{ekert} A.~K.~Ekert, Phys.\ Rev.\ Lett.\ {\bf 67}, 661 (1991).
\bibitem{chsh} J.~F.~Clauser, M.~A.~Horne, A.~Shimony, and R.~A.~Holt, Phys.\ Rev.\ Lett.\ {\bf 23}, 880 (1969).
\bibitem{ch}J.~F.~Clauser and M.~A.~Horne, Phys.\ Rev.\ D {\bf 10}, 526 (1974).
\bibitem{kaszlikowski} D.~Kaszlikowski, P.~Gnacinski, M.~Zukowski, W.~Miklaszewski, and A.~Zeilinger, Phys.\ Rev.\ Lett.\ {\bf 85}, 4418 (200).
\bibitem{collins} D.~Collins, N.~Gisin, N.~Linden, S.~Massar, and S.~Popescu, Phys.\ Rev.\ Lett.\ {\bf 88}, 040404 (2002).
\bibitem{son} W.~Son, Jinhyoung Lee, and M.~S.~Kim, Phys.\ Rev.\ Lett.\ {\bf 96}, 060406 (2006).
\bibitem{gisin} N.~Gisin, quant-ph/0702021 (2007).
\bibitem{cleve} R.~Cleve, P.~Hoyer, B.~Toner, and J.~Watrous, Proceedings of the 19th IEEE Annual Conference on Computational Complexity 2004 Amherst, MA , (IEEE Conference Proceedings, New York, 2004), pp.\ 236-249 and quant-ph/040407.
\bibitem{cabello} M.~Sadiq, P.~Badziag, M.~Bourennane, and A.~Cabello, Phys.\ Rev.\ A {\bf 87}, 012128 (2013) and A.~Cabello, S.~Severini, and A.~Winter, arXiv:1010.2163.
\bibitem{abramsky} S.~Abarmsky and L.~Hardy, Phys.\ Rev.\ A {\bf 85} 062114 (2012).
\bibitem{cirelson} B.~S.~Cirel'son, Lett.\ Math.\ Phys.\ {\bf 4}, 93 (1980).
\bibitem{wehner} S.~Wehner, Phys.\ Rev.\ A {\bf 73}, 022110 (2006). t
\bibitem{pal} K.~Pal and T.~Vertesi, Phys.\ Rev.\ A {\bf 79}, 022120 (2009).
\bibitem{werner} R.~F.~Werner and M.~M.~Wolf, Phys.\ Rev.\ A {\bf 64}, 032112 (2001).
\bibitem{jordan} P.~X.~Gallagher and R.~J.~Proulx, in \emph{Contributions to Algebra}, H.~Bass, P.~Cassidy, and J.~Kovacic, eds.\ (Academic Press, New York, 1977).  
\end{thebibliography}
\end{document}